\documentclass{iopart}
\usepackage{iopams}

\newtheorem{Proposition}{Proposition}
\newtheorem{Lemma}{Lemma}

\newtheorem{Definition}{Definition}

\newenvironment{proof}{\noindent{\bf proof.}}{\hspace{\stretch{1}}\opensquare}

\newcommand{\Real}{{\textrm{Re}}}

\begin{document}

\title{Gluing of completely positive maps}
\author{Johan \AA berg}
\address{Department of Quantum Chemistry, Uppsala University, 
Box 518, SE-751 20 Uppsala, Sweden}
\ead{johan.aaberg@kvac.uu.se}

\begin{abstract}
Gluings of completely positive maps (CPMs) are defined and
investigated. As a brief description of this concept consider a pair
of `evolution machines', each with the ability to evolve the internal
state of a `particle' inserted into its input. Each of these machines
is characterized by a channel describing the operation the internal
state has experienced when the particle is returned at the
output. Suppose a particle is put in a superposition between the input
of the first and the second machine. Here it is shown that the total
evolution caused by a pair of such devices is not uniquely determined
by the channels of the two machines. Such `global' channels describing
the machine pair are examples of \emph{gluings} of the two single
machine channels. Under the limiting assumption that all involved
Hilbert spaces are finite-dimensional, an expression which generates
all \emph{subspace preserving} gluings of a given pair of CPMs, is
derived. The nature of the non-uniqueness of gluings and its relation
to a proposed definition of \emph{subspace locality}, is discussed.
\end{abstract}

\pacs{03.65.-w, 03.67.-a}

\section{Introduction}
\label{se1}
Completely positive maps (CPMs) and trace preserving completely
positive maps are useful tools to describe operations in quantum
mechanics.  This investigation appears in a family of papers
\cite{ref2}, \cite{ref1} devoted to the study of completely positive
maps with respect to properties tied to orthogonal sum decomposition
of the Hilbert spaces of quantum systems. In \cite{ref1} the concept
of \emph{subspace preserving} CPMs is introduced. In \cite{ref2} a
definition of \emph{subspace locality} is proposed, while here the
concept of gluings of CPMs is introduced.

To give an intuitive picture of the concept of gluing, imagine an
apparatus which evolves the state of quantum systems. Imagine a
`particle' of some kind and a machine constructed such that when the
particle is inserted in the input of the machine, the internal state
of the particle is evolved and the particle is returned at the
output. The operation of the machine is modeled by a `channel', i.e. a
trace preserving CPM.  Imagine now that we have two such machines,
each characterized by a channel and suppose we have one single
particle. The question is: suppose this single particle is put in
superposition between the two inputs of the two evolution machines,
what would be the output, when the two machines act on this
superposition?  At first glance it may seem as if the evolution of the
superposition is trivially determined by the two channels
characterizing the action of the two machines. This is however not the
case, as shall be demonstrated here. The two channels do not provide
sufficient information to uniquely determine the total evolution.

To rephrase the problem in more mathematical terms, there are two
Hilbert spaces $\mathcal{H}_{1}$ and $\mathcal{H}_{2}$, the first
representing the pure input states of machine $1$, the second
representing pure input states of machine $2$. The total input state
Hilbert space $\mathcal{H}$ can be described as the orthogonal sum of
the two separate state spaces $\mathcal{H} =
\mathcal{H}_{1}\oplus\mathcal{H}_{2}$. For the quite general type of
evolution we are considering here, allowing interaction with ancillary
quantum systems, it is necessary to use density operators to describe
states of quantum systems. A channel maps density operators to density
operators. The channel $\Phi_{1}$ characterizing machine $1$ maps
density operators on $\mathcal{H}_{1}$ to density operators on
$\mathcal{H}_{1}$. Likewise the channel $\Phi_{2}$ of machine $2$ maps
density operators on $\mathcal{H}_{2}$ to density operators on
$\mathcal{H}_{2}$. The operation caused by the two machines acting in
combination on one single input, can be described as a channel which
maps density operators on $\mathcal{H}$ to density operators on
$\mathcal{H}$. To rephrase the above question: does $\Phi_{1}$ and
$\Phi_{2}$ determine $\Phi$ uniquely? The negative answer to this
question implies that there exist several channels $\Phi$ which, in
some sense, are `compatible' with the two channels $\Phi_{1}$ and
$\Phi_{2}$.  We call such a channel $\Phi$ a \emph{gluing} of
$\Phi_{1}$ and $\Phi_{2}$. One of the purposes of this investigation
is to find an explicit expression for describing all possible gluings
of given channels $\Phi_{1}$ and $\Phi_{2}$. This is achieved by first
proving that all trace preserving gluings of channels have to be
\emph{subspace preserving} \cite{ref1} and using tools developed in
\cite{ref1}.

One further question is: what happens if one imposes the restriction
that the two evolution machines should act independently of each
other? Hence, there should be no interaction, communication, or
sharing of resources like entangled or correlated quantum systems,
which would enable a correlated action. This restriction is added by
using a definition of \emph{subspace local} channels suggested in
\cite{ref2}. The set of gluings which satisfies this additional
condition is deduced. One may perhaps imagine that the non-uniqueness
of the gluings has its cause in the freedom of the machines to
interact (or to share correlated systems), and that if all such
`dependencies' are cut away, the non-uniqueness of the gluings would
disappear. When the gluings are restricted to be subspace local, the
non-uniqueness get reduced, but some non-trivial non-uniqueness
actually does remain. Hence, even if the machines are acting
independently, the CPMs of the two machines are still not sufficient
to determine the joint action of the two devices. This is resolved by
noting that the channel $\Phi_{1}$ for device $1$ (and $\Phi_{2}$ for
device $2$) is actually not a full description of the action of this
machine, in this context. By providing the `missing parts', the
gluings are uniquely determined.

The derivations performed here are made under the limiting assumption
that all Hilbert spaces appearing are finite-dimensional.  The author
believes that much of the material derived here do have
generalizations, with suitable technical modifications, to separable
Hilbert spaces. This question will, however, not be treated here.

The structure of this article is the following. In section \ref{se3}
gluings of CPMs is introduced. Explicit expressions to generate all
possible gluings of two given CPMs are deduced. In section \ref{se4} we
turn to the special case of SP gluings which fulfills the additional
condition of being subspace local. In section \ref{se5} the theory is
illustrated with some simple examples of gluings. In section \ref{nu}
we discuss some conceptual aspects of the non-uniqueness of gluings
and its relation to subspace locality. A summary is presented in
section \ref{se6}.
\section{Gluings}
\label{se3}
We begin by introducing some notation, terminology and conventions to
be used throughout this article.

$\mathcal{H}$ (with various subscripts) denotes a finite-dimensional
complex Hilbert space. Completely positive maps (CPMs) take trace
class operators of one Hilbert space to trace-class operators on
another Hilbert space.  On a finite-dimensional Hilbert space the set
of trace class operators coincides with the set of linear
operators. Since this study is restricted to finite-dimensional
Hilbert spaces we let CPMs operate on the set of linear operators on
the Hilbert space in question.

The set of linear operators on $\mathcal{H}$ is denoted
$\mathcal{L}(\mathcal{H})$. The set of linear operators from
$\mathcal{H}_{S}$ to $\mathcal{H}_{T}$ is denoted
$\mathcal{L}(\mathcal{H}_{S},\mathcal{H}_{T})$.

If a CPM $\phi$ maps elements in $\mathcal{L}(\mathcal{H}_{S})$ to
elements in $\mathcal{L}(\mathcal{H}_{T})$, we say that
$\mathcal{H}_{S}$ is the \emph{source space} (or just \emph{source})
of $\phi$, and that $\mathcal{H}_{T}$ is the \emph{target space} (or
just \emph{target}) of $\phi$.  When discussing a CPM the spaces
$\mathcal{H}_{S}$ and $\mathcal{H}_{T}$ are always assumed to be the
source and the target space of the CPM in question, unless otherwise
stated.

CPMs can always be constructed via Kraus representations
\cite{Kraus}. Given a CPM $\phi$ there exists some set of operators
$\{V_{k}\}_{k}\subset\mathcal{L}(\mathcal{H}_{S},\mathcal{H}_{T})$
such that $\phi(Q)= \sum_{k}V_{k}QV_{k}^{\dagger}$ for all
$Q\in\mathcal{L}(\mathcal{H}_{S})$. We refer to such a set
$\{V_{k}\}_{k}$ as a \emph{Kraus representation} of $\Phi$.

To every CPM there exists a linearly independent Kraus representation
(proposition 6 in \cite{ref1}). The number of elements in a linearly
independent Kraus representation of a CPM $\phi$ is called the
\emph{Kraus number} (proposition 6 in \cite{ref1}) of the CPM and is
denoted $K(\phi)$.

$\mathcal{H}_{s1}$ and $\mathcal{H}_{s2}$ denote subspaces of
$\mathcal{H}_{S}$. They are assumed to be orthogonal complements of
each other, such that $\mathcal{H}_{S} = \mathcal{H}_{s1}\oplus
\mathcal{H}_{s2}$ ($\oplus$ denotes orthogonal sum). Analogously
$\mathcal{H}_{T} = \mathcal{H}_{t1}\oplus \mathcal{H}_{t2}$. Finally
it is assumed that each of the subspaces $\mathcal{H}_{s1}$,
$\mathcal{H}_{s2}$, $\mathcal{H}_{t1}$, and $\mathcal{H}_{t2}$ are at
least one-dimensional. To all these subspaces are orthogonal
projectors associated. To $\mathcal{H}_{s1}$ belongs the projector
$P_{s1}$, to $\mathcal{H}_{t1}$ belongs $P_{t1}$, etc.

Given an operator $V:\mathcal{H}_{s1}\rightarrow \mathcal{H}_{t1}$, we
will in some expressions handle it as if it was an operator
$V':\mathcal{H}_{S}\rightarrow \mathcal{H}_{T}$, where $V'$ acts as
$V$ on $\mathcal{H}_{s1}$ and as the zero operator on
$\mathcal{H}_{s1}$ and is extended linearly to whole
$\mathcal{H}_{S}$. For notational simplicity we do not differ between
$V$ and $V'$. Another abuse of notation, in the same spirit as the
previous one, concerns CPMs. Given a CPM $\phi$ with source space
$\mathcal{H}_{s1}$ and target space $\mathcal{H}_{t1}$, we will in
some expressions handle it as if it had source space $\mathcal{H}_{S}$
and target $\mathcal{H}_{T}$. If $\phi$ has a Kraus representation
$\{V_{k}\}_{k}$, then this `extended' CPM can be constructed as
$\{V'_{k}\}_{k}$ with $V'_{k}$ as described above. We will not make
any difference between these CPMs. It is to be noted that if $\Phi$ is
a trace preserving CPM with source $\mathcal{H}_{s1}$ and target
$\mathcal{H}_{t1}$, then it is not trace preserving if regarded as
having source $\mathcal{H}_{S}$ and target $\mathcal{H}_{T}$.

Let $\phi$ be a CPM with source space $\mathcal{H}_{S}$ and target
space $\mathcal{H}_{T}$ and let $\mathcal{H}_{s1}$ be a subspace of
$\mathcal{H}_{S}$. We let $\overline{\phi}$ be defined as the
restriction (in the ordinary sense) of $\phi$ to the subset
$\mathcal{L}(\mathcal{H}_{s1})$. We say that $\overline{\phi}$ is the
\emph{restriction in source space to} $\mathcal{H}_{s1}$. Since the
set of density operators on the subspace $\mathcal{H}_{s1}$ is a
subset of the set of density operators on $\mathcal{H}_{S}$, follows
that $\overline{\phi}$ is a positive map. Let $\mathcal{H}_{n}$ be
n-dimensional and let $I_{n}$ be the identity CPM with source and
target $\mathcal{H}_{n}$. That $\overline{\phi}$ is \emph{completely}
positive follows since $\overline{\phi}\otimes I_{n}$ is the
restriction in source space of $\phi\otimes I_{n}$ to
$\mathcal{H}_{s1}\otimes\mathcal{H}_{n}$. Above we concluded that the
restriction in source space of a positive map is positive, hence
$\overline{\phi}\otimes I_{n}$ is positive for each $n$. Hence, by
definition \cite{Kraus}, $\widetilde{\phi}$ is completely positive.

Next we define restriction in target space. Given a subspace
$\mathcal{H}_{t1}$ of the target space of a CPM $\phi$, we define the
mapping $\phi'(Q)= P_{t1}\phi(Q)P_{t1}$, $\forall
Q\in\tau(\mathcal{H}_{S})$.  In this equation another abuse of
notation has been made. $P_{t1}$ denotes a mapping from
$\mathcal{H}_{T}$ to $\mathcal{H}_{T}$, but here we rather intend
$P_{t1}$ to be regarded as a map from $\mathcal{H}_{T}$ to the
subspace $\mathcal{H}_{t1}$. Since the mapping $\eta(Q) = P_{t1}Q
P_{t1}$ is written in the Kraus representation form (no matter the
exact status of $P_{t1}$) it follows that $\eta$ is a CPM. Hence,
$\phi'$ is a composition of two CPMs and hence is a CPM. We can
conclude:
\begin{Lemma}
\begin{itemize}
\item The restriction in source space of a CPM is a CPM. 
Moreover, the restriction in source space of a trace preserving CPM is
trace preserving.
\item The restriction in target space of a CPM is a CPM.
\end{itemize} 
\end{Lemma}  
Note that in difference with restriction in source space, a
restriction in target space of a trace preserving CPM is, in general,
\emph{not} trace preserving.
\begin{Definition}\rm
Let $\phi_{1}$ be a CPM with source $\mathcal{H}_{s1}$ and target
$\mathcal{H}_{t1}$ and let $\phi_{2}$ be a CPM with source
$\mathcal{H}_{s2}$ and target $\mathcal{H}_{t2}$. A CPM $\phi$ with
source $\mathcal{H}_{S}$ and target $\mathcal{H}_{T}$ is said to be a
\emph{gluing} of $\phi_{1}$ and $\phi_{2}$, if $\phi_{1}$ is the
result of restriction in target to $\mathcal{H}_{t1}$ and in source to
$\mathcal{H}_{s1}$ of $\phi$, and if $\phi_{2}$ is the result of
restriction in target to $\mathcal{H}_{t2}$ and in source to
$\mathcal{H}_{s2}$ of $\phi$. If moreover $\phi$ is \emph{subspace
preserving} (SP) from $(\mathcal{H}_{s1},\mathcal{H}_{s2})$ to
$(\mathcal{H}_{t1},\mathcal{H}_{t2})$ then we say that $\phi$ is an
\emph{SP gluing} of $\phi_{1}$ and $\phi_{2}$.
\end{Definition}
For the trace preserving CPMs an especially simple relation holds for
gluings and SP gluings
\begin{Proposition}
\label{ekviv}
Let $\Phi$ be a trace preserving CPM with source $\mathcal{H}_{S}$ and
target $\mathcal{H}_{T}$. $\Phi$ is SP from
$(\mathcal{H}_{s1},\mathcal{H}_{s2})$ to
$(\mathcal{H}_{t1},\mathcal{H}_{t2})$ if and only if $\Phi$ is a trace
preserving gluing of two trace preserving CPMs, $\Phi_{1}$ with source
$\mathcal{H}_{s1}$ and target $\mathcal{H}_{t1}$, and $\Phi_{2}$ with
source $\mathcal{H}_{s2}$ and target $\mathcal{H}_{t2}$.
\end{Proposition}

\begin{proof}
We begin to prove the ``if'' part of the proposition.  From $\Phi_{1}$
being trace preserving it follows that
$\Tr(P_{t1}\Phi(|\psi\rangle\langle\psi|)) =
\Tr(\Phi_{1}(|\psi\rangle\langle\psi|)) = 1$, for all normalized
$|\psi\rangle\in\mathcal{H}_{s1}$.  Since $\Phi$ is trace preserving,
it follows that $\Tr(P_{t2}\Phi(|\psi\rangle\langle\psi|)) = 0$, for
all $|\psi\rangle\in\mathcal{H}_{s1}$. By an analogous argument
follows $\Tr(P_{t1}\Phi(|\psi\rangle\langle\psi|)) = 0$, for all
$|\psi\rangle\in\mathcal{H}_{s2}$. From the two last equations
follows, by definition \cite{ref1}, that $\Phi$ is SP from
$(\mathcal{H}_{s1},\mathcal{H}_{s2})$ to
$(\mathcal{H}_{t1},\mathcal{H}_{t2})$.

The ``only if'' part follows from the fact that any SP CPM can be
decomposed as (see proposition 2 in \cite{ref1})
\begin{displaymath}
\fl \Phi(Q) = P_{t1}\Phi(P_{s1}QP_{s1})P_{t1} +
 P_{t1}\Phi(P_{s1}QP_{s2})P_{t2} + P_{t2}\Phi(P_{s2}QP_{s1})P_{t1} + 
P_{t2}\Phi(P_{s2}QP_{s2})P_{t2}.
\end{displaymath}
The CPM $P_{t1}\Phi(P_{s1}QP_{s1})P_{t1}$ is essentially (with a
purely technical modification of the source and target spaces) the
restriction in source to $\mathcal{H}_{s1}$ and in target to
$\mathcal{H}_{t1}$. Moreover, this restriction is trace preserving. An
analogous reasoning holds for
$P_{t2}\Phi(P_{s2}QP_{s2})P_{t2}$. Hence, $\Phi$ is a trace preserving
gluing of two channels.
\end{proof}

In \cite{ref1} an expression for the set of all SP CPMs has been
derived. This expression will be used to derive an expression for all
the SP gluings of two CPMs with known linearly independent Kraus
representations. For the sake of convenience proposition 10 in
\cite{ref1} is restated here
\begin{Proposition}
\label{alliss}
Let $\{V_{k}\}_{k=1}^{K}$ be a basis of
$\mathcal{L}(\mathcal{H}_{s1},\mathcal{H}_{t1})$, $K =
\dim\mathcal{H}_{s1}\dim\mathcal{H}_{t1}$, and let
$\{W_{l}\}_{l=1}^{L}$ be a basis of
$\mathcal{L}(\mathcal{H}_{s2},\mathcal{H}_{t2})$, $L =
\dim\mathcal{H}_{s2}\dim\mathcal{H}_{t2}$. The mapping $\phi$, defined
by
\begin{equation}
\label{allagl2}
\fl\phi(Q) =  \sum_{kk'}A_{k,k'}V_{k}QV_{k'}^{\dagger}+
\sum_{ll'}B_{l,l'}W_{l}QW_{l'}^{\dagger} + 
\sum_{kl}C_{kl}V_{k}QW_{l}^{\dagger} + 
\sum_{kl}C_{kl}^{*}W_{l}QV_{k}^{\dagger},
\end{equation}
for all $Q\in\mathcal{L}(\mathcal{H}_{S})$, is an SP CPM from
$(\mathcal{H}_{s1},\mathcal{H}_{s2})$ to
$(\mathcal{H}_{t1},\mathcal{H}_{t2})$ if and only if the matrices $A =
[A_{k,k'}]_{k,k'=1}^{K}$, $B = [B_{l,l'}]_{l,l'=1}^{L}$ and $C =
[C_{k,l}]_{k=1,l=1}^{K,L}$ fulfill the relations
\begin{equation}
\label{kondi0}
A\geq 0,\quad B\geq 0,
\end{equation}
\begin{equation}
\label{kondi}
P_{A,0}C = 0,\quad CP_{B,0} = 0,\quad A \geq CB^{\ominus}C^{\dagger},
\end{equation}
where $B^{\ominus}$ denotes the Moore-Penrose pseudo inverse
\cite{LanTis}, \cite{MPi}, \cite{MPi2} of $B$. $P_{A,0}$ denotes the
orthogonal projector onto the zero eigenspace of $A$ and analogously
for $P_{B,0}$.

Moreover, (\ref{allagl2}) defines a bijection between the set of all
CPMs which are SP from $(\mathcal{H}_{s1},\mathcal{H}_{s2})$ to
$(\mathcal{H}_{t1},\mathcal{H}_{t2})$, and the set of all triples of
matrices $A$,$B$,$C$ fulfilling the conditions (\ref{kondi0}) and
(\ref{kondi}).
\end{Proposition}
The proposition above gives an expression for all possible SP CPMs
with respect to given decompositions of the source and target
space. We are interested, not in all such CPMs, but only those SP CPMs
which are gluings of two given CPMs. Given the knowledge of linearly
independent Kraus representations of the CPMs $\phi_{1}$ and
$\phi_{2}$, it is possible to construct a rather compact expression
for the set of SP gluings.
\begin{Proposition}
\label{allglue}
Let $\phi_{1}$ be a CPM with source space $\mathcal{H}_{s1}$ and
target space $\mathcal{H}_{t1}$. Let $\phi_{2}$ be a CPM with source
space $\mathcal{H}_{s2}$ and target space $\mathcal{H}_{t2}$.  Let
$\{V_{n}\}_{n=1}^{N}$ be a linearly independent Kraus representation
\cite{ref1} of $\phi_{1}$ and let $\{W_{m}\}_{m=1}^{M}$ be a linearly
independent Kraus representation of $\phi_{2}$. Then $\phi$ is an SP
gluing of $\phi_{1}$ and $\phi_{2}$, if and only if $\phi$ can be
written
\begin{eqnarray}
\label{pr21}
\fl\phi(Q) =  \sum_{n=1}^{N}V_{n}QV_{n}^{\dagger} + 
\sum_{m=1}^{M}W_{m}QW_{m}^{\dagger}\nonumber\\ 
+\sum_{m =1,n=1}^{N,M} C_{n,m}V_{n}QW_{m}^{\dagger} + 
\sum_{m=1,n=1}^{N,M} C_{n,m}^{*}W_{m}QV_{n}^{\dagger},\quad\forall
Q\in\mathcal{L}(\mathcal{H}_{S}),
\end{eqnarray} 
where the matrix $C = [C_{n,m}]_{n=1,m=1}^{N,M}$ fulfills the condition
\begin{equation}
\label{kondvar1}
I_{N}\geq CC^{\dagger},
\end{equation}
where $I_{N}$ denotes the $N\times N$ identity matrix.  Moreover,
(\ref{pr21}) defines a bijection between the set of all SP gluings of
$\phi_{1}$ and $\phi_{2}$, and the set of all matrices $C$ fulfilling
condition (\ref{kondvar1}).
\end{Proposition}
The condition $I_{N}\geq CC^{\dagger}$ is equivalent to $I_{M}\geq
C^{\dagger}C$, with $I_{M}$ the $M\times M$ identity matrix.  These
conditions in turn are equivalent to the condition that the largest
singular value of $C$ should be less than or equal to one. These
comments can be derived by using singular value decomposition
\cite{LanTis} of $C$. (Use $C = U_{1}^{\dagger}\widetilde{C}U_{2}$
where these operators are defined as in the proof of proposition
\ref{KnSD}.)

The matrix $C$ in the above proposition we refer to as the \emph{gluing
matrix}. Note that the choices of linearly independent Kraus
representations are arbitrary. The gluing matrix depends on this choice, 
but not the set of gluings.

If $\phi_{1}$ and $\phi_{2}$ are trace preserving, then proposition
\ref{allglue} gives all the trace preserving gluings. This follows
from proposition \ref{ekviv}.

\begin{proof} 
We begin to prove that any CPM on the form (\ref{pr21}) is an SP
gluing of $\phi_{1}$ and $\phi_{2}$. One can check that any CPM which
can be written on the form (\ref{pr21}) has $\phi_{1}$ as restriction
in source to $\mathcal{H}_{s1}$ and in target to $\mathcal{H}_{t1}$,
and has $\phi_{2}$ as restriction in source to $\mathcal{H}_{s2}$ and
in target to $\mathcal{H}_{t2}$. Hence, $\phi$ is a gluing of
$\phi_{1}$ and $\phi_{2}$. It remains to show that every mapping
$\phi$ defined by (\ref{pr21}) is an SP CPM. (The main part is to
prove that $\phi$ is a CPM.) Complete the set $\{V_{n}\}_{n=1}^{N}$
into a basis $\{V_{k}\}_{k=1}^{K}$ of
$\mathcal{L}(\mathcal{H}_{s1},\mathcal{H}_{t1})$, in such a way that
the first $N$ elements form the linearly independent Kraus
representation $\{V_{n}\}_{n=1}^{N}$.  Define the matrix
$A=[A_{k,k'}]_{k,k'=1}^{K}$ by
\begin{equation}
\label{Amatrdef}
 A_{k,k'} =\delta_{k,k'},\quad\forall k,k'\leq N,\quad {\rm and} \quad
 A_{k,k'}=0 \quad {\rm else}.
\end{equation}
Clearly $\phi_{1}(Q) =
\sum_{k,k'=1}^{K}A_{k,k'}V_{k}QV_{k'}^{\dagger}$. Similarly, complete
the set $\{W_{m}\}_{m=1}^{M}$ into a basis of
$\mathcal{L}(\mathcal{H}_{s2},\mathcal{H}_{t2})$ and construct the
matrix $B$ similarly as $A$ was constructed. Let the matrix
$\widetilde{C}=[\tilde{c}_{k,l}]_{k=1,l=1}^{K,L}$ be defined as
$\tilde{c}_{k,l} = C_{k,l}$ for $k\leq N$ and $l\leq M$, and
$\tilde{c}_{k,l}=0$ else. One can verify that the triple
$(A,B,\widetilde{C})$ fulfills the conditions (\ref{kondi0}) and
(\ref{kondi}) of proposition \ref{alliss}. Moreover, $\phi$ is the
result when $(A,B,\widetilde{C})$ are used in (\ref{allagl2}). Hence,
according to proposition \ref{alliss}, $\phi$ is an SP CPM.

It is to be shown that every SP gluing of $\phi_{1}$ and $\phi_{2}$
can be written on the form (\ref{pr21}). Complete (like above) the
linearly independent Kraus representations $\{V_{n}\}_{n=1}^{N}$ and
$\{W_{m}\}_{m=1}^{M}$ into bases. Since we are searching for all SP
gluings of $\phi_{1}$ and $\phi_{2}$, all of them can be written
(since they are SP) on the form (\ref{allagl2}), with respect to the
above chosen bases, where each SP CPM corresponds to a triple
$(A,B,\widetilde{C})$. For such an SP CPM to be a gluing of $\phi_{1}$
and $\phi_{2}$, one can see that a necessary condition is that
$\phi_{1}(Q)= \sum_{k,k'=1}^{K}A_{k,k'}V_{k}QV_{k'}^{\dagger}$ and
$\phi_{2}(Q)= \sum_{l,l'=1}^{L}B_{l,l'}W_{l}QW_{l'}^{\dagger}$. Since
these matrices are uniquely determined by the choice of bases (see
proposition 5 in \cite{ref1}) it follows that the matrix $A
=[A_{k,k'}]_{k,k'=1}^{K}$ is the one defined in (\ref{Amatrdef}). A
similar reasoning holds for $B$. Using the conditions (\ref{kondi}) it
follows that only a sub-matrix of the matrix $\widetilde{C}$ is
non-zero. This sub-matrix is the sub-matrix defined by $C =
[\tilde{c}_{n,m}]_{n=1,m=1}^{N,M}$. Moreover one can check that the
conditions (\ref{kondi}) on $(A,B,\widetilde{C})$ imply that $C$
fulfills (\ref{kondvar1}).
 
That equation (\ref{pr21}) defines a bijection between the set of all
SP gluings of $\phi_{1}$, $\phi_{2}$ and the set of all matrices $C$
fulfilling condition (\ref{kondvar1}), follows from the bijectivity
stated in proposition \ref{alliss}.
\end{proof}

One may wonder how the gluing matrix $C$ of proposition \ref{allglue} 
changes if one makes other choices of linearly independent Kraus
representations of $\phi_{1}$ and $\phi_{2}$. From proposition 7 in
\cite{ref1} it is known that there is a bijective correspondence
between the set of $K(\phi)\times K(\phi)$ unitary matrices and the
set of linearly independent Kraus representations of $\phi$. From
proposition 7 in \cite{ref1}, it follows that if the linearly
independent Kraus representations are changed, then there exists a
unitary $K(\phi_{1})\times K(\phi_{1})$ matrix $U_{1}$, and a unitary
$K(\phi_{2})\times K(\phi_{2})$ matrix $U_{2}$, such that the new
gluing matrix $C'$ relates to the old as $C' =
U_{1}CU_{2}^{\dagger}$. One may note that this implies that the set of
singular values of the gluing matrix is independent of the choice of
linearly independent Kraus representations.

The following proposition shows that the Kraus number \cite{ref1} of
an SP gluing, is constrained by the Kraus numbers of the CPMs which
are glued.
\begin{Proposition}
\label{KnSD}
If the CPM $\phi$ is an SP gluing of the CPMs $\phi_{1}$ and $\phi_{2}$, then
\begin{equation}
\max(K(\phi_{1}),K(\phi_{2})) \leq K(\phi)\leq K(\phi_{1})+K(\phi_{2}).
\end{equation} 
The left equality holds if and only if the gluing matrix $C$ given by
proposition \ref{allglue}, has $\min(K(\phi_{1}),K(\phi_{2}))$
singular values with value $1$, counted with multiplicity. Moreover,
$K(\phi_{1})+K(\phi_{2})- K(\phi)$ is the number of singular values
with value $1$ of the matrix $C$, counted with multiplicity.
\end{Proposition}

\begin{proof}
Let $K_{1}=K(\phi_{1})$ and $K_{2}=K(\phi_{2})$.  By proposition 6 in
\cite{ref1} the Kraus number of a CPM $\phi$ is equal to the number
of non-zero eigenvalues of a representation matrix $F$, given by
proposition 5 in \cite{ref1}. This is true regardless of the choice
of basis of $\mathcal{L}(\mathcal{H}_{T},\mathcal{H}_{S})$. Given
linearly independent Kraus representations of $\phi_{1}$ and
$\phi_{2}$, the same construction as in the proof of proposition
\ref{allglue} can be made. Hence, we complete the two sets of linearly
independent Kraus representations to become bases. With respect to
this choice of bases we find that the representation matrix only has a
sub-matrix which is non-zero. This sub-matrix has the form
\begin{displaymath}
F = \left[\begin{array}{cc}
 I_{1} & C\\ C^{\dagger} & I_{2} \end{array} \right],
\end{displaymath}
where $I_{1}$ denotes the $K_{1}\times K_{1}$ identity matrix, and
$I_{2}$ the $K_{2}\times K_{2}$ identity matrix. The number of
non-zero eigenvalues of the total representation matrix is the number
of non-zero eigenvalues of the sub-matrix $F$.  Without loss of
generality we may assume $K_{1}\leq K_{2}$. The matrix $C$ can be
transformed into an especially simple form by applying a singular
value decomposition \cite{LanTis}. There exists a $K_{1}\times K_{1}$
unitary matrix $U_{1}$ and a $K_{2}\times K_{2}$ unitary matrix
$U_{2}$ such that $U_{1}CU_{2}^{\dagger} = \widetilde{C}$, where
$\widetilde{C}$ is a $K_{1}\times K_{2}$ matrix which is composed from
a diagonal $K_{1}\times K_{1}$ matrix, with the singular values as
diagonal elements, and a $K_{1}\times (K_{2}-K_{1})$ zero
matrix. (Singular values are always non-negative.)
\begin{equation}
\widetilde{C} = \left[\begin{array}{cccccc} r_{1}& & &0&\ldots &0\\
&\ddots & & \vdots & & \vdots\\
& & r_{K_{1}}& 0 &\ldots & 0 \end{array}\right].
\end{equation}
Let
\begin{equation} 
\widetilde{F} = (U_{1}\oplus U_{2})
F(U_{1}^{\dagger}\oplus U_{2}^{\dagger})= 
\left[\begin{array}{cc}
I_{1} & \widetilde{C}\\
\widetilde{C}^{\dagger} & I_{2}
\end{array}\right].
\end{equation}
 Since $\widetilde{F}$ is obtained from $F$ by a unitary
 transformation, both have the same set of eigenvalues. One can check
 that the unitary matrix
\begin{equation}
\widetilde{U} = 
\left[\begin{array}{ccc} \frac{1}{\sqrt{2}}I_{K_{1}} &
 \frac{1}{\sqrt{2}}I_{K_{1}} & 0\\
\frac{1}{\sqrt{2}}I_{K_{1}} & -\frac{1}{\sqrt{2}}I_{K_{1}} & 0 \\
0 & 0 & I_{K_{2}-K_{1}}
 \end{array}\right]
\end{equation}
diagonalizes $\widetilde{F}$, in such a way that the first $K_{1}$
eigenvalues are $1+r_{k}$ for $k=1,\ldots, K_{1}$, the next $K_{1}$
eigenvalues are $1-r_{k}$ for $k=1,\ldots, K_{1}$, and the remaining
$K_{2}-K_{1}$ eigenvalues are all $1$.  Since $r_{k}\geq 0$, the
number of non-zero eigenvalues of $\widetilde{F}$ is $K_{1}+K_{2}-N$,
where $N$ is the number of singular values with value $1$. Since the
number of non-zero singular values of $C$ maximally can be
$\min(K_{1},K_{2})$, and since $K_{1}+K_{2}-\min(K_{1},K_{2})=
\max(K_{1},K_{2})$, the proposition follows.
\end{proof}

The set of SP gluings of two given CPMs forms a convex set. For the
rest of this section we find out the extreme points of this set.
Let $I(N,M)$ denote the set of all complex $N\times M$ matrices $C$
such that $CC^{\dagger}\leq I_{N}$. For the sake of simplicity it is
assumed that $N\leq M$.  Let $EI(N,M)$ denote the set of complex
$N\times M$ matrices with $CC^{\dagger}= I_{N}$.
\begin{Lemma}
\label{partissub}
$I(N,M)$ is a convex set.
$EI(N,M)$ is the set of extreme points of $I(N,M)$.
\end{Lemma}
A complex $N\times M$ matrix belongs to $EI(N,M)$ if and only it has
precisely $N$ non-zero singular values, all of value $1$ (if $N\leq
M$).

\begin{proof}
First note that the lemma is true if $N=1$. Hence, we may in the
following assume $N\geq 2$.

We have to prove that every element of $I(N,M)$ can be formed as a
convex combination of elements in $EI(N,M)$.  Let
$\{\overline{c}_{j}\}_{j=1}^{N}$ be an orthonormal basis of
$\mathbb{C}^{N}$ and let $\{\overline{d}_{k}\}_{k=1}^{N}$ be an
orthonormal set in $\mathbb{C}^{M}$. (Regard $\overline{c}_{j}$ and
$\overline{d}_{k}$ as column vectors.) Let $D^{(0)} =
\sum_{k=1}^{N}\overline{c}_{k}\overline{d}_{k}^{\dagger}$. For $1\leq
n\leq N-1$, let $D^{(n)}=
-\sum_{k=1}^{n}\overline{c}_{k}\overline{d}_{k}^{\dagger} +
\sum_{k=n+1}^{N}\overline{c}_{k}\overline{d}_{k}^{\dagger}$. Let
$D^{(N)}= - D^{(0)}$. All the matrices $D^{(n)}$, $n=0,\ldots,N$
belongs to $EI(N,M)$.  Define
$H^{(n)}=\frac{1}{2}D^{(0)}+\frac{1}{2}D^{(n)}$, for
$n=0,\ldots,N$. By construction each $H^{(n)}$ is in the convex hull
of $EI(N,M)$.

Let $C$ be an arbitrary $N\times M$ complex matrix. As such it can be
decomposed using a singular value decomposition \cite{LanTis}. There
exist non-negative real numbers $r_{n}$, $n=1,\ldots,N$, an
orthonormal basis $\{\overline{c}_{k}\}_{k=1}^{N}$ of
$\mathbb{C}^{N}$, and an orthonormal set
$\{\overline{d}_{k}\}_{k=1}^{N}$ in $\mathbb{C}^{M}$, such that $C =
\sum_{k=1}^{N}r_{k}\overline{c}_{k}\overline{d}_{k}^{\dagger}$.  (Here
the assumption $N\leq M$ is used.)  One can check that $C\in I(N,M)$
if and only if $r_{k}\leq 1$ for all $k=1,\ldots,N$. Assume $C\in
I(N,M)$. Without loss of generality we may assume that the orthonormal
sets are ordered in such a way that $r_{1}\leq r_{2}\leq \ldots \leq
r_{N}$.  Let $\lambda_{0} = r_{1}$ and $\lambda_{n} = r_{n+1}-r_{n}$
for $n=1,\ldots,N-1$. Let $\lambda_{N} = 1-r_{N}$. By construction it
follows that $\lambda_{n}\geq 0$. Moreover, $\sum_{n=0}^{N}\lambda_{n}
= 1$. One can check that $\sum_{n=0}^{N}\lambda_{n}H^{(n)} = C$.
Hence, $C$ is in the convex hull of $EI(N,M)$.

Now it is to be proved that no element in $EI(N,M)$ can be written as
a non-trivial convex combination of two elements in $I(N,M)$. Suppose
$C\in EI(N,M)$ is such that there exists $C_{1},C_{2}\in I(M,N)$ and
$0<p<1$, such that $C = pC_{1} + (1-p)C_{2}$. Then,
\begin{equation}
\label{eqett}
\fl I_{N} = CC^{\dagger} = 
(pC_{1} + (1-p)C_{2})(pC_{1} + (1-p)C_{2})^{\dagger},
\end{equation} 
since $C\in EI(N,M)$. Moreover, since $C_{1},C_{2}\in I(N,M)$ it
follows that
\begin{eqnarray}
\label{eqtvaa}
\fl & (pC_{1} + (1-p)C_{2})(pC_{1} + (1-p)C_{2})^{\dagger}\\
&\leq p^{2}I_{N} + (1-p)^{2}I_{N} + p(1-p)(C_{2}C_{1}^{\dagger}+
C_{1}C_{2}^{\dagger}).
\end{eqnarray}
By combining (\ref{eqett}) and (\ref{eqtvaa}) and using $0<p<1$ one
finds that 
\begin{equation}
\label{huvforml}
2I_{N}\leq C_{2}C_{1}^{\dagger}+C_{1}C_{2}^{\dagger}.
\end{equation} 
 If $\overline{a}$ is an arbitrary normalized element of $\mathbb{C}^{N}$,
then it follows from (\ref{huvforml}) that 
\begin{equation}
\label{realoverl}
1\leq \Real(\overline{a}^{\dagger}C_{1}C_{2}^{\dagger}\overline{a}).
\end{equation} 
From $C_{1},C_{2}\in I(N,M)$ and $||\overline{a}||=1$ it follows that  
$||C_{1}^{\dagger}\overline{a}||\leq 1$ and
$||C_{2}^{\dagger}\overline{a}||\leq 1$. 
The two last inequalities together with (\ref{realoverl}) 
can be true only if 
$C_{1}^{\dagger}\overline{a}=C_{2}^{\dagger}\overline{a}$. Since
$\overline{a}$ is an arbitrary normalized vector it follows that
$C_{1}=C_{2}$. Hence, $C = pC_{1} + (1-p)C_{2}=C_{1}=C_{2}$. This is a
trivial convex combination. Hence, there is no element in $EI(N,M)$
which is a non-trivial convex combination of elements in $I(N,M)$.
\end{proof}

\begin{Proposition}
The set of all SP gluings of two given CPMs $\phi_{1}$ and $\phi_{2}$
is convex. An SP gluing $\phi$ in this set is an extreme point if and
only if the matrix $C$ of proposition \ref{allglue} has precisely
$\min(K(\phi_{1}),K(\phi_{2}))$ singular values with value $1$, which
occurs if and only if $K(\phi) = \max(K(\phi_{1}),K(\phi_{2}))$.
\end{Proposition}
\begin{proof}
Without loss of generality we may assume $K(\phi_{1})\leq
K(\phi_{2})$. The set of gluing matrices of proposition \ref{allglue}, 
with respect to some arbitrary choices of linearly independent Kraus
representations of $\phi_{1}$ and $\phi_{2}$, is
$I(K(\phi_{1}),K(\phi_{2}))$. According to lemma \ref{partissub} this
is a convex set.  From lemma \ref{partissub} we know that the set of
extreme points is $EI(K(\phi_{1}),K(\phi_{2}))$. A matrix $C$ is an
element of $EI(K(\phi_{1}),K(\phi_{2}))$ if and only if $C$ has
precisely $K(\phi_{1})$ non-zero singular values, all with value $1$. 
Since we have assumed $K(\phi_{1})\leq K(\phi_{2})$, the general 
condition is that there should be $\min(K(\phi_{1}),K(\phi_{2}))$ 
singular values,
all with value $1$. From proposition \ref{KnSD} we know that this
occurs if and only if $K(\phi) = \max(K(\phi_{1}),K(\phi_{2}))$.
\end{proof}

\section{Local subspace preserving gluings}
\label{se4}
In \cite{ref2} the concepts of subspace local (SL) channels and local
subspace preserving (LSP) channels, are introduced. Here these are
used to construct gluings which are subspace local. In terms of the
`evolution machines' this corresponds to a combination of two machines
which do not interact or share any correlated systems. This is an
attempt to formalize the concept of a combination of two independently
acting devices. LSP gluing are defined as follows:
\begin{Definition}\rm
If $\Phi_{1}$ is a channel with source $\mathcal{H}_{s1}$ and target
$\mathcal{H}_{t1}$, and if $\Phi_{2}$ is a channel with source
$\mathcal{H}_{s2}$ and target $\mathcal{H}_{t2}$ then a trace
preserving gluing of $\Phi_{1}$ and $\Phi_{2}$ is called a \emph{LSP}
gluing if $\Phi$ is SL from $(\mathcal{H}_{s1},\mathcal{H}_{s2})$ to
$(\mathcal{H}_{t1},\mathcal{H}_{t2})$.
\end{Definition}
The reason why these gluings are called ``LSP gluings'' and not ``SL
gluings'' is that these automatically are LSP channels. This follows
from the fact that trace preserving gluings of channels are SP gluings
(proposition \ref{ekviv}), and that the intersection between the set
of SP channels and the set of subspace local channels, is the set of
LSP channels (see proposition 5 in \cite{ref2}).

By combining proposition \ref{allglue}, proposition \ref{ekviv} and
proposition 2 in \cite{ref2} the following proposition is obtained.
\begin{Proposition}
\label{islglue}
Let $\Phi_{1}$ be a trace preserving CPM with source space
$\mathcal{H}_{s1}$ and target space $\mathcal{H}_{t1}$. Let $\Phi_{2}$
be trace preserving CPM with with source space $\mathcal{H}_{s2}$ and
target space $\mathcal{H}_{t2}$.  Let $\{V_{n}\}_{n=1}^{N}$ be a
linearly independent Kraus representation of $\Phi_{1}$. Let
$\{W_{m}\}_{m=1}^{M}$ be a linearly independent Kraus representation
of $\Phi_{2}$. Then $\Phi$ is an LSP gluing of $\Phi_{1}$ and
$\Phi_{2}$ if and only if $\Phi$ can be written
\begin{equation}
\label{pr}
\fl \Phi(Q)= \sum_{n=1}^{N}V_{n}QV_{n}^{\dagger} + 
\sum_{m=1}^{M}W_{m}QW_{m}^{\dagger}+ VQW^{\dagger} + 
WQV^{\dagger} ,\quad\forall Q\in\mathcal{L}(\mathcal{H}_{S}),
\end{equation}
with $V = \sum_{n=1}^{N} c_{1,n}V_{n}$, $W = \sum_{m
=1}^{M}c_{2,m}W_{m}$, where the vectors $c_{1} = [c_{1,n}]_{n=1}^{N}$
and $c_{2} = [c_{2,m}]_{m=1}^{M}$ fulfill the conditions $||c_{1}||^{2}
= \sum_{n=1}^{N}|c_{1,n}|^{2} \leq 1$ and $||c_{2}||^{2} =
\sum_{m=1}^{M}|c_{2,m}|^{2} \leq 1$.
\end{Proposition}
\begin{Proposition}
\label{singvalcond}
A trace preserving gluing of two channels is an LSP gluing if and only
if the matrix $C$ given by proposition \ref{allglue}, has at most one
non-zero singular value, counted with multiplicity.
\end{Proposition}
One can note that the vectors $c_{1}$ and $c_{2}$ are not uniquely
determined by the LSP gluing. However, the gluing matrix $C_{nm} =
c_{1,n}c_{2,m}^{*}$ is. The same non-uniqueness affects the operators
$V$ and $W$. However, the non-uniqueness is rather `mild', since it is
essentially limited to variations of a single complex number.

\begin{proof}
We begin with the ``only if'' part. By comparing proposition
\ref{islglue} and \ref{allglue}, one can see that if $c_{1} =
[c_{1,n}]_{n=1}^{N}$ and $c_{2} = [c_{2,m}]_{m=1}^{M}$ fulfill the
conditions of proposition \ref{islglue}, then the matrix $C$ of
proposition \ref{allglue} is $C =c_{1}c_{2}^{\dagger}$. ($c_{1}$ and
$c_{2}$ are regarded as column vectors.) Since a matrix on this form
has at most one non-zero singular value counted with multiplicity, the
``only if'' part follows.

For the ``if'' part, let $C$ be the matrix resulting from proposition
\ref{allglue}. If $C$ has at most one non-zero singular value (counted
with multiplicity), then there exist vectors $\widetilde{c}_{1}$ and
$\widetilde{c}_{2}$ such that $C$ can be written $C =
\widetilde{c}_{1}\widetilde{c}_{2}^{\dagger}$. In case $C=0$ the
condition for being an LSP gluing is clearly fulfilled. Hence, without
loss of generality we may assume $C$ is not the zero matrix. Hence,
$\widetilde{c}_{1}$ and $\widetilde{c}_{2}$ must be non-zero.  The
matrix $C$ fulfills $CC^{\dagger}\leq I_{N}$. This condition is, for
the special form of $C$ considered here, translated into
$||\widetilde{c}_{1}||^{2}\,||\widetilde{c}_{2}||^{2}\leq 1$. Let
$c_{1} = \widetilde{c}_{1}
(||\widetilde{c}_{2}||/||\widetilde{c}_{1}||)^{1/2}$ and $c_{2} =
\widetilde{c}_{2}
(||\widetilde{c}_{1}||/||\widetilde{c}_{2}||)^{1/2}$, then $c_{1}$ and
$c_{2}$ fulfill the conditions of proposition \ref{islglue}.
\end{proof}
\begin{Proposition}
\label{condisl}
Let $\Phi_{1}$ be a trace preserving CPM and let $\Phi_{2}$ be a trace
preserving CPM such that $K(\Phi_{2}) = 1$. Every trace preserving
gluing of $\Phi_{1}$ and $\Phi_{2}$ is an LSP gluing.
\end{Proposition}
As a direct corollary of this proposition it follows that every trace
preserving gluing of a trace preserving CPM and an identity CPM is an
LSP gluing.

\begin{proof}
The matrix $C$ of proposition \ref{allglue} is a $K(\Phi_{1})\times
K(\Phi_{2})$ matrix. Since $K(\Phi_{2}) = 1$ this matrix can have at
most one non-zero singular value counted with multiplicity. Hence, by
proposition \ref{singvalcond}, the the statement of the proposition
follows.
\end{proof}

Being an LSP gluing puts a very stringent condition on the relation
between the Kraus numbers of the gluing and the glued channels, as the
following proposition shows. This proposition follows directly by
combining proposition \ref{KnSD} and proposition \ref{singvalcond}.
\begin{Proposition}
\label{islkrcond}
If a trace preserving CPM $\Phi$ is an LSP gluing of the trace
preserving CPMs $\Phi_{1}$ and $\Phi_{2}$, then $K(\Phi_{1}) +
K(\Phi_{2})-1\leq K(\Phi)\leq K(\Phi_{1}) + K(\Phi_{2})$.
\end{Proposition}
\section{Some illustrations}
\label{se5}
In this section the theory is illustrated with some simple
examples. In addition to serving as illustration, some of these
derivations also indicate possible directions for future studies. We
attempt to compare the ability of LSP and SP gluings to preserve
superposition, in some sense.

A quantum channel is called unitary on $\mathcal{H}_{s1}$ if it can be
written $\Phi_{1}(Q) = U_{1}QU_{1}^{\dagger}$, where $U_{1}$ is
unitary operator on $\mathcal{H}_{s1}$. Given a unitary channel
$\Phi_{1}$ on $\mathcal{H}_{s1}$ and a unitary channel $\Phi_{2}$ on
$\mathcal{H}_{s2}$, what is the set of trace preserving gluings of
these channels? Both the channels have Kraus number $1$. Hence, all
gluings of them have to be LSP (proposition \ref{condisl}). By
proposition \ref{allglue} all those gluings can be written
\begin{equation}
\label{uglue}
\Phi(Q) = U_{1}QU_{1}^{\dagger} +  U_{2}QU_{2}^{\dagger} + 
cU_{1}QU_{2}^{\dagger} +  c^{*}U_{2}QU_{1}^{\dagger},
\end{equation}
where $|c|\leq 1$. Using a polar decomposition of $c$ into $c =
re^{i\theta}$, (\ref{uglue}) can be rewritten as
\begin{equation}
\label{uglueomsk}
\fl \Phi(Q) = (1-r)U_{1}QU_{1}^{\dagger} + (1-r)U_{2}QU_{2}^{\dagger}
 + r(U_{1}+e^{-i\theta}U_{2})Q(U_{1}+e^{-i\theta}U_{2})^{\dagger}.
\end{equation}
It is straightforward to check that $U_{1}+e^{-i\theta}U_{2}$ is a
unitary operator. Let us focus on the two extreme cases $r=0$ and
$r=1$ to see what happens with an initial superposition of states
localized in the two subspaces. In case $r=1$, $\Phi_{r=1}$ is unitary
and hence maps pure states to pure states. In some sense these
channels preserve the superposition between the two subspaces. To see
this, let $|\psi\rangle\in\mathcal{H}_{S}$ be normalized but else
arbitrary. It can be written $|\psi\rangle =
\alpha_{1}|\psi_{1}\rangle + \alpha_{2}|\psi_{2}\rangle$, where
$|\psi_{1}\rangle$ is some normalized state in $\mathcal{H}_{s1}$ and
$|\psi_{2}\rangle$ some normalized state in $\mathcal{H}_{s2}$.  Then
$\Phi(|\psi\rangle\langle\psi|)= |\psi'\rangle\langle\psi'|$, where
$|\psi'\rangle = \alpha_{1}U_{1}|\psi_{1}\rangle +
\alpha_{2}e^{-i\theta}U_{2}|\psi_{2}\rangle$. Hence, the weights
$|\alpha_{1}|^{2}$ and $|\alpha_{2}|^{2}$ of the two subspaces in the
superposition, have not changed under the mapping. In some sense the
`amount' of superposition is preserved under these types of channels.
In case $r=0$, the channel $\Phi_{r=0}$ completely destroys the
superposition, in the sense that any superposition between states
localized in the two subspaces is turned into a mixture of states
localized in the two subspaces. $\Phi_{r=0}(|\psi\rangle\langle\psi|)
= |\alpha_{1}|^{2}U_{1}|\psi_{1}\rangle\langle
\psi_{1}|U_{1}^{\dagger} +
|\alpha_{2}|^{2}U_{2}|\psi_{2}\rangle\langle
\psi_{2}|U_{2}^{\dagger}$.  Hence, in one extreme $r=1$ superposition
is preserved, and in the other extreme $r=0$ superposition is
completely destroyed. From (\ref{uglueomsk}) follows that intermediate
choices of $r$ give a partial destruction of superposition.  It has to
be emphasized that all these channels, no matter the choice of $r$,
are possible to perform subspace locally. Hence, in this specific
case, the question of whether the channel preserve superposition or
not, is independent of whether the channel is possible to perform
subspace locally or not. We will see however, that in the case of
other more complicated gluings, LSP gluings seems to be less able to
preserve superposition, compared to general SP gluings.

It is to be noted that a channel do not need to be implemented
subspace locally just because it is subspace local. The subspace
locality only states that it \emph{can} be implemented subspace
locally. An example is $\Phi_{r}$ with $0<r<1$. Since $\Phi_{r} =
(1-r)\Phi_{r=0}+r\Phi_{r=1}$, and since $r$ can be interpreted as a
probability, follows that $\Phi_{r}$ can be implemented by using a
random generator choosing the channel $\Phi_{r=0}$ with probability
$r$ and channel $\Phi_{r=1}$ with probability $(1-r)$. This is not a
local implementation since the outcome of the random generator has to
be distributed to both locations. However, this channel is subspace
local, as have been demonstrated above.

For comparison we here consider a channel which is SP but not LSP. Let
the unitary channels $\Phi_{a}$ and $\Phi_{b}$ be defined by
\begin{eqnarray}
\label{ex}
\Phi_{a}(Q)=U_{a}QU_{a}^{\dagger},\quad \Phi_{b}(Q)=
U_{b}QU_{b}^{\dagger},\nonumber\\
U_{a} = V_{s1} + P_{s2},\quad U_{b} = P_{s1} + V_{s2},
\end{eqnarray}
where $V_{s1}$ and $V_{s2}$ satisfy, $V_{s1} V_{s1}^{\dagger} =
V_{s1}^{\dagger}V_{s1} =P_{s1}$ and $V_{s2} V_{s2}^{\dagger} =
V_{s2}^{\dagger}V_{s2} =P_{s2}$. Moreover, we assume that
$\{V_{s1},P_{s1}\}$ is a linearly independent set, and that $\{V_{s2},
P_{s2}\}$ is a linearly independent set. (Both $\Phi_{a}$ and
$\Phi_{b}$ are LSP gluings of the type considered in the previous
example.) Consider the convex combination $\Phi =
\frac{1}{2}\Phi_{a}+\frac{1}{2}\Phi_{b}$. The mapping $\Phi$ can be
realized with a random generator determining which of the operations
$\Phi_{a}$ or $\Phi_{b}$ is performed, each with probability one half.
Hence, with probability one half a unitary channel is operating
locally in subspace $\mathcal{H}_{s1}$ and the identity CPM acts on
$\mathcal{H}_{s2}$. With probability one half the opposite
happens. Because of the construction with the shared outcome of the
random generator, this implementation is not subspace local. As the
previous example has shown, the channel may still be subspace local.
However, it is possible to prove that $\Phi$ is not an LSP
channel. The channel $\Phi = \frac{1}{2}\Phi_{a}+\frac{1}{2}\Phi_{b}$
can be written
\begin{equation}
\label{andr}
\fl\Phi(Q) = \Phi_{1}(Q) + \Phi_{2}(Q) + \frac{1}{2}V_{s1}QP_{s2} + 
\frac{1}{2}P_{s1}QV_{s2}^{\dagger} + \frac{1}{2}V_{s2}QP_{s1} + 
\frac{1}{2}P_{s2}QV_{s1}^{\dagger},
\end{equation} 
where $\Phi_{1}$ and $\Phi_{2}$ are given by $\Phi_{1}(Q) =
\frac{1}{2}V_{s1}Q V_{s1}^{\dagger} +\frac{1}{2}P_{s1}Q P_{s1}$ and
$\Phi_{2}(Q) = \frac{1}{2}V_{s2}Q V_{s2}^{\dagger} +\frac{1}{2}P_{s2}Q
P_{s2}$. $\Phi_{1}$ can be regarded as a trace preserving CPM with
source and target $\mathcal{H}_{s1}$ (but it is not trace preserving
if regarded as a CPM with source and target $\mathcal{H}_{S}$.) 
Likewise $\Phi_{2}$ is trace preserving with source and target
$\mathcal{H}_{s2}$. The set
$\{\frac{1}{\sqrt{2}}V_{s1},\frac{1}{\sqrt{2}}P_{s1}\}$ is a linearly
independent Kraus representation of $\Phi_{1}$, and
$\{\frac{1}{\sqrt{2}}V_{s2},\frac{1}{\sqrt{2}}P_{s2}\}$ is a linearly
independent Kraus representation of $\Phi_{2}$.  Equation (\ref{andr})
is on the form required by proposition \ref{allglue}, with matrix $C =
\left[ \begin{array}{cc} 0 & 1\\ 1 & 0 \end{array}\right]$. This
matrix has two non-zero singular values (both are $1$). By proposition
\ref{singvalcond} it follows that $\Phi$ cannot be an LSP gluing. We
conclude that $\Phi$ is an SP gluing, but not an LSP gluing, of the
two channels $\Phi_{1}$ and $\Phi_{2}$. Hence, $\Phi$ is not subspace
local.

A very simple type of channel maps every state (pure or mixed) to a
fixed pure state. Let $|\psi_{1}\rangle\in\mathcal{H}_{s1}$ be
normalized. Define $\Phi_{1}$ to be the channel which maps any density
operator on $\mathcal{H}_{s1}$ to the density operator
$|\psi_{1}\rangle\langle\psi_{1}|$. If $\{|s1_{k}\rangle\}_{k}$ is an
arbitrary orthonormal basis of $\mathcal{H}_{s1}$, then a linearly
independent Kraus representation of $\Phi_{1}$ is
$\{|\psi_{1}\rangle\langle s1_{k}|\}_{k}$. Likewise a similar channel
$\Phi_{2}$, with source and target $\mathcal{H}_{s2}$, which maps all
states to a pure state $|\psi_{2}\rangle\in\mathcal{H}_{s2}$, has a
linearly independent Kraus representation $\{|\psi_{2}\rangle\langle
s2_{l}|\}_{l}$. All trace preserving gluings of $\Phi_{1}$ and
$\Phi_{2}$ can be written
\begin{eqnarray}
\fl\Phi(Q) = |\psi_{1}\rangle\langle\psi_{1}|\Tr(P_{s1}Q) + 
|\psi_{2}\rangle\langle\psi_{2}|\Tr(P_{s2}Q)\nonumber\\
  +|\psi_{1}\rangle\langle\psi_{2}|\sum_{kl}C_{kl}\langle
  s1_{k}|Q|s2_{l}\rangle
  +|\psi_{2}\rangle\langle\psi_{1}|\sum_{kl}C_{kl}^{*}\langle
  s2_{l}|Q|s1_{k}\rangle,
\end{eqnarray}
for all $Q\in\mathcal{L}(\mathcal{H}_{S})$.  In the special case of
LSP gluings, $\Phi$ takes the form $\Phi(Q) =
|\psi_{1}\rangle\langle\psi_{1}|\Tr(P_{s1}Q) +
|\psi_{2}\rangle\langle\psi_{2}|\Tr(P_{s2}Q) +
|\psi_{1}\rangle\langle\psi_{2}|\langle a|Q|b\rangle +
|\psi_{2}\rangle\langle\psi_{1}|\langle b|Q|a\rangle$, where
$|a\rangle\in\mathcal{H}_{s1}$ and $|b\rangle\in\mathcal{H}_{s2}$
fulfill $||a||\leq 1$ and $||b||\leq 1$, but are otherwise
arbitrary. If we choose $|a\rangle=|\psi_{1}\rangle$ and
$|b\rangle=|\psi_{2}\rangle$, the result is a channel of the form
\begin{eqnarray}
\label{spec}
\fl \Phi(Q)=  |\psi_{1}\rangle\langle\psi_{1}|\Tr(P_{s1}Q) + 
|\psi_{2}\rangle\langle\psi_{2}|\Tr(P_{s2}Q)\nonumber\\ 
  +|\psi_{1}\rangle\langle\psi_{2}|\langle \psi_{1}|Q|\psi_{2}\rangle
  + |\psi_{2}\rangle\langle\psi_{1}|\langle \psi_{2}|Q|\psi_{1}\rangle
  ,\quad\forall Q\in\mathcal{L}(\mathcal{H}_{S}).
\end{eqnarray}
Any pure state on the form $|\psi\rangle = \alpha|\psi_{1}\rangle +
\beta|\psi_{2}\rangle$ is left intact by channel (\ref{spec}). Hence,
in this case the superposition between the two subspaces is
preserved. Let $|\psi'\rangle = \alpha|\psi_{1}^{\perp}\rangle +
\beta|\psi_{2}^{\perp}\rangle$, where $|\psi_{1}^{\perp}\rangle$ is
any state in $\mathcal{H}_{s1}$ which is orthogonal to
$|\psi_{1}\rangle$, and analogously for
$|\psi_{2}^{\perp}\rangle$. (Hence, $\mathcal{H}_{s1}$ and
$\mathcal{H}_{s2}$ are at least two-dimensional.) It follows that
$\Phi(|\psi'\rangle\langle\psi'|)=
|\alpha|^{2}|\psi_{1}\rangle\langle\psi_{1}|+
|\beta|^{2}|\psi_{2}\rangle\langle\psi_{2}|$. In this case the
superposition is destroyed and leaves a mixture of states localized in
the two subspaces. Hence, this channel preserve the superposition only
for a quite limited set of input states. Now we turn to the more
general SP gluings. Assume
$\dim(\mathcal{H}_{s1})=\dim(\mathcal{H}_{s2})$. Then it is possible
to choose $C_{kl} = \delta_{kl}$. This gluing is such that any pure
input state on the form $\alpha|s1_{k}\rangle +\beta|s2_{k}\rangle$ is
mapped to the pure state $\alpha|\psi_{1}\rangle +
\beta|\psi_{2}\rangle$. Hence, in general the states will not be
preserved, but in some sense the `amount' of superposition is
preserved. In this case a number of families of input states are
mapped in a way that 'preserves' the superposition, in contrast with
the case of LSP gluings, where there is only one such family. This
indicates that SP gluings have better abilities to preserve
superposition than do LSP gluings.

Finally we consider two rather simple channels to see under what
conditions these are SP or LSP.  Let $\mathcal{H}_{a}$ be some
arbitrary finite-dimensional Hilbert space and let $\sigma$ be some
density operator on $\mathcal{H}_{a}$. Let $\rho$ denote density
operators on a space $\mathcal{H}_{S}$, and define the channel
$\Lambda(\rho)= \rho\otimes\sigma$. Hence, $\Lambda$ has source space
$\mathcal{H}_{S}$ and target space
$\mathcal{H}_{S}\otimes\mathcal{H}_{a}$. Given an arbitrary
decomposition $\mathcal{H}_{S}=\mathcal{H}_{s1}\oplus\mathcal{H}_{s2}$
($\mathcal{H}_{s1}$ and $\mathcal{H}_{s2}$ are at least
one-dimensional), is $\Lambda$ SP from
$(\mathcal{H}_{s1},\mathcal{H}_{s2})$ to
$(\mathcal{H}_{s1}\otimes\mathcal{H}_{a},
\mathcal{H}_{s2}\otimes\mathcal{H}_{a})$? 
Moreover, is it LSP?  The channel $\Lambda$ is SP since
$\Tr(P_{s1}\otimes\hat{1}_{a}\Lambda(Q)) = \Tr(P_{s1}Q\otimes\sigma) =
\Tr(P_{s1}Q)$, for any $Q\in\mathcal{L}(\mathcal{H}_{S})$. Hence, by
proposition 4 in \cite{ref1} $\Lambda$ is an SP channel.  We wish to
find out whether $\Lambda$ is an LSP channel, or not. Since $\Lambda$
is SP, it has to be a gluing of two trace preserving CPMs. One may
verify that these two are $\Psi_{1}(Q) = P_{s1}QP_{s1}\otimes\sigma$
and $\Psi_{2}(Q) = P_{s2}QP_{s2}\otimes\sigma$. Let
$\{\lambda_{k}\}_{k=1}^{K}$ denote the \emph{non-zero} eigenvalues and
$\{|\lambda_{k}\rangle\}_{k=1}^{K}$ a corresponding orthonormal set of
eigenvectors of $\sigma$. In a slightly odd notation
$\{\sqrt{\lambda_{k}}|\lambda_{k}\rangle P_{s1}\}_{k=1}^{K}$ is a
linearly independent Kraus representation of $\Phi_{1}$, and
$\{\sqrt{\lambda_{k}}|\lambda_{k}\rangle P_{s2}\}_{k=1}^{K}$ of
$\Phi_{2}$. (A more strict notation would be
$\{\sum_{l}\sqrt{\lambda_{k}}|\lambda_{k}\rangle|s1_{l}\rangle\langle
s1_{1}|\}_{k=1}^{K}$ for some orthonormal basis
$\{|s1_{l}\rangle\}_{l}$ of $\mathcal{H}_{s1}$.) When applying
proposition \ref{allglue} to $\Phi$, with these choices of linearly
independent Kraus representations, it is found that $C = I_{K}$, where
$I_{K}$ denotes the $K\times K$ identity matrix. Hence, by using
proposition \ref{singvalcond} it is found that $\Phi$ can be LSP from
$(\mathcal{H}_{s1},\mathcal{H}_{s2})$ to
$(\mathcal{H}_{s1}\otimes\mathcal{H}_{a},
\mathcal{H}_{s2}\otimes\mathcal{H}_{a})$,
if and only if $K=1$. Hence, $\Phi$ is LSP if and only if the density
operator $\sigma$ represents a pure state.

The, in a sense, `opposite' channel to $\Lambda$ is the partial
trace. $\Tr_{a}$ is SP from
$(\mathcal{H}_{s1}\otimes\mathcal{H}_{a},
\mathcal{H}_{s2}\otimes\mathcal{H}_{a})$
to $(\mathcal{H}_{s1},\mathcal{H}_{s2})$, since $\Tr(P_{s1}\Tr_{a}(Q))
= \Tr(P_{s1}\otimes\hat{1}_{a}Q)$ for all
$Q\in\mathcal{L}(\mathcal{H}_{S}\otimes\mathcal{H}_{a})$. Again we
refer to proposition 4 of \cite{ref1} to conclude that partial trace
is SP. As such it is a trace preserving gluing of two trace preserving
CPMs, one with linearly independent Kraus representation
$\{P_{s1}\langle a_{k}|\}_{k=1}^{K}$ and one with linearly independent
Kraus representation $\{P_{s2}\langle a_{k}|\}_{k=1}^{K}$, where
$\{|a_{k}\rangle\}_{k=1}^{K}$ is an arbitrary orthonormal basis of
$\mathcal{H}_{a}$. The matrix $C$ given by proposition \ref{allglue},
with respect to the chosen linearly independent Kraus representations,
is $C =I_{K}$. Hence, the partial trace is LSP if and only if
$\mathcal{H}_{a}$ is one-dimensional. We can conclude the following.
\begin{Proposition}
\label{pTr}
Let $\mathcal{H}_{a}$ be finite-dimensional and let $\sigma$ be a
density operator on $\mathcal{H}_{a}$.
\begin{itemize}
\item $\Lambda(Q) = Q\otimes\sigma,\quad
\forall Q\in\mathcal{L}(\mathcal{H}_{S})$ 
is an SP channel from $(\mathcal{H}_{s1},\mathcal{H}_{s2})$ to
$(\mathcal{H}_{s1}\otimes\mathcal{H}_{a},\mathcal{H}_{s2}
\otimes\mathcal{H}_{a})$. It is LSP from 
$(\mathcal{H}_{s1},\mathcal{H}_{s2})$ to
$(\mathcal{H}_{s1}\otimes\mathcal{H}_{a},\mathcal{H}_{s2}
\otimes\mathcal{H}_{a})$
if and only if $\sigma$ is pure.
\item $\Tr_{a}$ is an SP channel from 
$(\mathcal{H}_{s1}\otimes\mathcal{H}_{a},
\mathcal{H}_{s2}\otimes\mathcal{H}_{a})$ to  
$(\mathcal{H}_{s1},\mathcal{H}_{s2})$ . It is LSP  from 
$(\mathcal{H}_{s1}\otimes\mathcal{H}_{a},
\mathcal{H}_{s2}\otimes\mathcal{H}_{a})$ to  
$(\mathcal{H}_{s1},\mathcal{H}_{s2})$ if and only if $\dim(\mathcal{H}_{a})=1$.
\end{itemize}
\end{Proposition}
Except for providing some examples of subspace non-local channels, the
channels of proposition \ref{pTr} can be used as building blocks to
construct the set of SP channels from the set of LSP
channels. Proposition \ref{SPrep1} and \ref{SPrep2} both show that
every SP channel can be constructed from LSP channels and a final
partial trace. The reason why both have been included is that while
proposition \ref{SPrep1} highlights the role of the partial trace,
proposition \ref{SPrep2} is perhaps more intuitively accessible, since
it constructs the channel as a more clearcut action on a
system-ancilla decomposition.
\begin{Proposition}
\label{SPrep1}
A channel $\Phi$ is SP from $(\mathcal{H}_{s1},\mathcal{H}_{s2})$ to
$(\mathcal{H}_{t1},\mathcal{H}_{t2})$ if and only if there exists a
Hilbert space $\mathcal{H}_{a}$ and a channel $\Psi$, which is LSP
from $(\mathcal{H}_{s1},\mathcal{H}_{s2})$ to
$(\mathcal{H}_{t1}\otimes\mathcal{H}_{a},
\mathcal{H}_{t2}\otimes\mathcal{H}_{a})$
and such that
\begin{equation}
\label{phirep1}
\Phi(Q) = \Tr_{a}\Psi(Q), \quad\forall Q\in\mathcal{L}(\mathcal{H}_{S}).
\end{equation}
\end{Proposition}
\begin{proof}
Since $\Phi$ is a composition of two SP channels, the ``if'' part
follows. To prove the ``only if'' part, we use that every Kraus
representation of an SP CPM can be written $\{V_{1,k}+V_{2,k}\}_{k}$,
where $P_{t1}V_{1,k}P_{s1} = V_{1,k}$ and $P_{t2}V_{2,k}P_{s2} =
V_{2,k}$ (proposition 1 in \cite{ref1}). This Kraus representation can
be chosen to be finite if the source and target space are
finite-dimensional (see proposition 6 in \cite{ref1}), which they are
by assumption. If the Kraus representation contains $K$ elements, let
$\mathcal{H}_{a}$ be a K-dimensional Hilbert space and let
$\{|a_{k}\rangle\}_{k=1}^{K}$ be an arbitrary orthonormal basis of
$\mathcal{H}_{a}$. Let the operators $V_{1}$ and $V_{2}$ be defined by
$V_{1}=\sum_{k=1}^{K}|a_{k}\rangle V_{1,k}$ and
$V_{2}=\sum_{k=1}^{K}|a_{k}\rangle V_{2,k}$. One can show that the CPM
$\Psi$ defined by the Kraus representation $\{V_{1}+V_{2}\}$ is trace
preserving and fulfills equation (\ref{phirep1}). $\Psi$ is a gluing of
a trace preserving CPM with linearly independent Kraus representation
$\{V_{1}\}$ (with source $\mathcal{H}_{s1}$ and target
$\mathcal{H}_{t1}\otimes\mathcal{H}_{a}$) and a trace preserving CPM
with linearly independent Kraus representation $\{V_{2}\}$. Since
$\Psi$ is a trace preserving gluing of of CPMs with Kraus number $1$,
it follows (proposition \ref{condisl}) that $\Psi$ is LSP.
\end{proof}
\begin{Proposition}
\label{SPrep2}
A channel $\Phi$ is SP from $(\mathcal{H}_{s1},\mathcal{H}_{s2})$ to
$(\mathcal{H}_{t1},\mathcal{H}_{t2})$ if and only if there exists a
Hilbert space $\mathcal{H}_{a}$, a normalized state
$|a\rangle\in\mathcal{H}_{a}$, and a channel $\Psi'$, which is LSP
from
$(\mathcal{H}_{s1}\otimes\mathcal{H}_{a},
\mathcal{H}_{s2}\otimes\mathcal{H}_{a})$ to
$(\mathcal{H}_{t1}\otimes\mathcal{H}_{a},
\mathcal{H}_{t2}\otimes\mathcal{H}_{a})$
and such that
\begin{equation}
\label{phirep2}
\Phi(Q) = \Tr_{a}\Psi'(Q\otimes|a\rangle\langle a|), \quad
\forall Q\in\mathcal{L}(\mathcal{H}_{S}).
\end{equation}
\end{Proposition}
\begin{proof}
The ``if'' part follows as in the proof of proposition
\ref{SPrep1}. For the ``only if'' part, let $\mathcal{H}_{a}$,
$\{|a_{k}\rangle\}_{k=1}^{K}$, $V_{1}$, and $V_{2}$ be as in in the
proof of proposition \ref{SPrep1}. The set $\{V_{1}\langle
a_{l}|\}_{l=1}^{K}$ is a linearly independent Kraus representation of
a trace preserving CPM with source
$\mathcal{H}_{s1}\otimes\mathcal{H}_{a}$ and target
$\mathcal{H}_{t1}\otimes\mathcal{H}_{a}$. Likewise $\{V_{2}\langle
a_{l'}|\}_{l'=1}^{K}$ is a linearly independent Kraus representation
of a trace preserving CPM with source
$\mathcal{H}_{s2}\otimes\mathcal{H}_{a}$ and target
$\mathcal{H}_{t2}\otimes\mathcal{H}_{a}$. We construct $\Psi'$ as a
trace preserving gluing of these two channels. Let
$|a\rangle\in\mathcal{H}_{a}$ be an arbitrary normalized state. Define
$\Psi'$ by $\Psi'(Q) = \sum_{l=1}^{K} V_{1}\langle a_{l}| Q
|a_{l}\rangle V_{1}^{\dagger} + \sum_{l'=1}^{K} V_{2}\langle a_{l'}| Q
|a_{l'}\rangle V_{2}^{\dagger} +\\ + V_{1}\langle a| Q |a\rangle
V_{2}^{\dagger} + V_{2}\langle a| Q |a\rangle V_{1}^{\dagger}$, for
all $Q\in\mathcal{L}(\mathcal{H}_{S}\otimes\mathcal{H}_{a})$.  Since
$|a\rangle = \sum_{k}c_{k}|a_{k}\rangle$ for some complex numbers
$(c_{k})_{k=1}^{K}$ such that $\sum_{k=1}|c_{k}|^{2} =1$, it follows
by proposition \ref{islglue} that $\Psi'$ is an LSP gluing from
$(\mathcal{H}_{s1}\otimes\mathcal{H}_{a},
\mathcal{H}_{s2}\otimes\mathcal{H}_{a})$ to
$(\mathcal{H}_{t1}\otimes\mathcal{H}_{a},
\mathcal{H}_{t2}\otimes\mathcal{H}_{a})$. One
can check that $\Psi'$ fulfills equation (\ref{phirep2}).
\end{proof}
\section{The non-uniqueness of gluings}
\label{nu}
The fact that gluings are not unique indicates that there is some
aspect of the joint evolution which is not captured by the two
channels alone. If we return to the picture of two evolution machines
and a single particle, it means that although we know precisely how
each of the machines alone handles a particle, that knowledge is not
enough to deduce how the two machines act jointly on a
superposition. In the case of SP gluings this is perhaps not very
surprising, since the SP gluing allows the machines to interact with
each other, or share some correlated resources like entangled pairs of
particles. If we accept the definition of subspace locality put
forward in \cite{ref2}, the two machines in an LSP gluing should
truly be independent of each other.  It may seem a reasonable guess
that the LSP gluing should be uniquely determined by the channels
$\Phi_{1}$ and $\Phi_{2}$.  However, by comparison of propositions
\ref{allglue} and \ref{islglue}, one finds that the set of gluings is
reduced when the assumption of subspace locality is added, but there
is still non-trivial non-uniqueness.  To understand the remaining
non-uniqueness we change perspective on this problem.

The definition of subspace locality is based on a second quantization
of the Hilbert space of the system, as described in \cite{ref2}. To
simplify the discussion we, instead of LSP gluings of two arbitrary
channels, concentrate on gluings of a channel and an identity
channel. The identity channel may correspond to a machine which does
nothing with the particle. We moreover assume identical source and
target spaces, and identical decompositions of these. Hence, we
consider gluings of the channel $\Phi_{1}$ with source and target
$\mathcal{H}_{1}$, and the identity channel $I_{2}$ with source and
target $\mathcal{H}_{2}$. The relevant part of the occupation state
space of the input of the first device is essentially
$\widetilde{\mathcal{H}}_{1}=
\mathcal{H}_{1}\oplus\textrm{Sp}\{|0_{1}\rangle\}$. The
state $|0_{1}\rangle$ is the 'vacuum state', with no particle
present. Hence, we may have single particle states, the vacuum state,
as well as various linear combinations of these, as input states. The
same construction is made for the input of the other machine. The
total Hilbert space is
$\widetilde{\mathcal{H}}_{1}\otimes\widetilde{\mathcal{H}}_{2}$. By
construction of subspace locality \cite{ref2} there corresponds a
channel $\widetilde{\Phi}_{1}\otimes\widetilde{I}_{2}$ to every gluing
of the channels $\Phi_{1}$ and $I_{2}$. The channel
$\widetilde{\Phi}_{1}$ has source and target
$\widetilde{\mathcal{H}}_{1}$, and $\widetilde{I}_{2}$ is the identity
channel with source and target $\widetilde{\mathcal{H}}_{2}$. The
channel $\widetilde{\Phi}_{1}\otimes\widetilde{I}_{2}$ describes the
action on the occupation state. If $\Phi_{1}$ has linearly independent
Kraus representation $\{V_{k}\}_{k}$, it can be shown that
$\widetilde{\Phi}_{1}$ has to be on the form
\begin{eqnarray}
\label{device}
\widetilde{\Phi}_{1}(Q) =  &\sum_{k}V_{k}QV_{k}^{\dagger} + 
|0\rangle\langle 0|Q|0\rangle\langle 0|\nonumber \\
 &+VQ|0\rangle\langle 0| + |0\rangle\langle 0|QV^{\dagger}
 ,\quad\forall Q\in\mathcal{L}(\mathcal{H}_{s1}\oplus
 \textrm{Sp}\{|0\rangle\}),
\end{eqnarray}  
where the operator $V$ is 
\begin{equation}
V = \sum_{k}c_{k}V_{k},
\end{equation}
where the complex numbers $c_{k}$ fulfill $\sum_{k}|c_{k}|^{2}\leq
1$. (Equation (\ref{device}) is equation (6) in the proof of
proposition 2 in \cite{ref2}.)  In words, $\widetilde{\Phi}_{1}$ is a
trace preserving gluing of $\Phi_{1}$ and a CPM which maps the vacuum
state to the vacuum state. Hence, the operator $V$ results from the
gluing of these two channels, which explains the form of $V$.

In view of equation (\ref{device}) one can see what is missing in the
description of the device. The channel $\Phi_{1}$ only describes what
happens when there is a particle `fully present' in the input of the
non-trivial device. The missing part is provided by the operator $V$,
which describes what happens to linear combinations of single particle
states and the vacuum state. Hence, a complete description of the
machine in this context is the channel $\widetilde{\Phi}_{1}$. This
channel can equivalently be described as the pair $(\Phi_{1},V)$. In
an LSP gluing (see proposition \ref{islglue}), two such pairs, one for
each of the two machines, \emph{uniquely} determine the gluing. This
reflects the fact that when an evolution device acts on a
superposition, there is, in some sense, an additional `degree of
freedom' involved, namely the presence non-presence of the
particle. The action on this additional degree of freedom has to be
specified. When this is done, the LSP gluing is uniquely
determined. This restores the intuitive notion that the joint action
of two independently acting devices should be possible to describe
using the knowledge of the action of the two devices alone, and
explains the non-uniqueness of LSP gluings. For more examples and
discussions, in a more specific context, the reader is referred to
\cite{refinterf}, where these matters are discussed in terms of
single-particle two-path interferometry.
\section{Summary}
\label{se6}
The concept of gluing of completely positive maps is introduced. To
give a brief description of this concept, consider a quantum system
with corresponding Hilbert space $\mathcal{H}$. This state space is
decomposed in an orthogonal sum of two subspaces
$\mathcal{H}=\mathcal{H}_{1}\oplus\mathcal{H}_{2}$. Operations on the
state of the quantum system is described by trace preserving CPMs
(channels), which map density operators on $\mathcal{H}$ to density
operators on $\mathcal{H}$.  Suppose channel $\Phi$ is such that when
an input state is localized in $\mathcal{H}_{1}$, the returned state
is again localized in subspace $\mathcal{H}_{1}$. By `localized' is
intended that $P_{1}\rho P_{1}= \rho$ where $\rho$ is the density
operator and $P_{1}$ is the projection operator onto
$\mathcal{H}_{1}$. Assume that this restricted mapping can be
described by the channel $\Phi_{1}$. Likewise if the initial state is
localized in subspace $\mathcal{H}_{2}$, the returned state is
localized in $\mathcal{H}_{2}$, and this mapping is described by the
channel $\Phi_{2}$. The 'global' channel $\Phi$ is an example of a
gluing of the two channels $\Phi_{1}$ and $\Phi_{1}$. The channel
$\Phi$ is not uniquely determined by $\Phi_{1}$ and $\Phi_{2}$. As
shown here, there exist several possible trace preserving gluings of
$\Phi_{1}$ and $\Phi_{2}$.

An expression is derived (proposition \ref{allglue}), by which it is
possible to generate all subspace preserving (SP) \cite{ref1} gluings
of two given CPMs, with respect to arbitrary linearly independent
Kraus representations \cite{ref1} of these. From this follows the
construction of all trace preserving gluings of given pairs of
channels.

Using a proposed definition of subspace locality of quantum operations
\cite{ref2}, it is possible to express all trace preserving gluings
of trace preserving CPMs which also fulfill the property of being
subspace local (proposition \ref{islglue}). Intuitively an operation
which is subspace local \cite{ref2} can be regarded as being caused
by two independent evolution machines. It is intended that these two
devices do not interact or share any correlated resources like
entangled pairs of particles. The fact that gluings are not uniquely
determined by the glued channels, is discussed. We focus on the
difference between the general trace preserving gluings of channels
and the subspace local gluings. It is shown that even if the gluing is
subspace local, the gluings are still not uniquely determined. It is
argued that this non-uniqueness is due to that the channels $\Phi_{1}$
and $\Phi_{2}$, given as characterizations of the two machines, are
not full descriptions of these two devices, in this context. It is
suggested that a more complete description of each of the two machines
is given by a pair $(\Phi_{1},V)$ where $V$ is a linear map from the
source space of $\Phi_{1}$ to its target space. It is shown that a
subspace local gluing is characterized by two such pairs, one
for each device. Except for providing insight in the nature of
gluings, this discussion also serves as a conceptual illustration of
the results obtained in \cite{ref2}. The developed theory is
illustrated with some examples of gluings.

\ack I thank Erik Sj\"oqvist for many  valuable comments and discussions
 on the manuscript. I also thank Marie Ericsson for discussions which 
started the train of thoughts leading to this investigation.
 Finally I thank Osvaldo Goscinski for reading and commenting the text.

\end{document}